\documentclass[10pt,conference]{IEEEtran}
\IEEEoverridecommandlockouts

\usepackage{amsmath}
\usepackage{graphicx}
\usepackage{booktabs}
\usepackage{balance}
\usepackage{xcolor}
\usepackage{subfigure}
\usepackage{xspace}
\usepackage{dsfont}
\newcommand{\blue}[1]{\textcolor{black}{#1}}
\newcommand{\purple}[1]{\textcolor{black}{#1}}

\AtBeginDocument{%
  \providecommand\BibTeX{{%
    \normalfont B\kern-0.5em{\scshape i\kern-0.25em b}\kern-0.8em\TeX}}}
\newcommand{\toolname} {{\sc ALPACAS}\xspace}

\begin{document}

\title{Simulation-Driven Automated End-to-End Test and Oracle Inference\thanks{The primary author is Shreshth Tuli who is a PhD student at Imperial College London. Shreshth implemented the work reported in this paper, while undertaking an internship at Meta platforms Inc. Remaining authors, who were also involved in supporting and guiding the internship, development of the technique and writing this paper are listed in alphabetical order. }}

\author{
\IEEEauthorblockN{Shreshth Tuli, Kinga Bojarczuk, Natalija Gucevska, Mark Harman, Xiao-Yu Wang, Graham Wright}
\IEEEauthorblockA{Meta Platforms Inc.}
}

\maketitle


\begin{abstract}
This \blue{ is the first work to report on inferential testing at scale in industry. Specifically, it reports the experience of automated testing of integrity systems at Meta. 
We built an internal tool called} \toolname for automated inference of end-to-end integrity tests. 
Integrity tests are designed to keep users safe online by checking that interventions take place when harmful behaviour occurs on a platform.
\toolname~infers not only the test input, but also the oracle, by observing production interventions to prevent harmful behaviour. 
This approach  allows Meta to automate the process of generating integrity tests for its platforms, such as Facebook and Instagram, which consist of hundreds of millions of lines of production code. We outline the design and deployment of \toolname, and report results for its coverage, number of tests produced at each stage of the test inference process, and their pass rates. 
Specifically, we demonstrate that using \toolname \purple{ significantly improves coverage from a manual test design for the particular aspect of integrity end-to-end testing it was applied to. }
Further, from a pool of 3 million data points, \toolname automatically yields 39 production-ready end-to-end integrity tests.
We also report that the \toolname-inferred test suite enjoys exceptionally low flakiness for end-to-end testing with its average in-production pass rate of 99.84\%. 
\end{abstract}


\section{Introduction}

Test automation remains a significant open challenge for the software industry \cite{garousi:exploring}.
Fully automated testing requires the automated {\em design} and execution of tests, and not merely their execution.
The oracle problem, \textit{i.e.}, automatically determining acceptable output for a given input, has remained a formidable barrier to test design automation for five decades.
As a result, there has been much research on automatically inferring partial oracles and/or oracle properties to ameliorate this problem.
Nevertheless, until now, there has been no report on the industrial deployment of oracle inference at scale.

In this paper we \blue{present} and evaluate Meta's automated testing tool,  \toolname: \underline{A}utomatically-\underline{L}earnt \underline{P}roduction-\underline{AC}cumulated \underline{A}ction \underline{S}imulations, that infers oracles automatically, at scale, from production traffic, on Meta Platforms such as Facebook and Instagram. 
\toolname is part of the Metamorphic Interaction Automaton (MIA)~\cite{jaetal:mia}, deployed on Meta's WW simulation platform~\cite{jaetal:ease-keynote}.
The simulation framework, WW,  enables us to perform multiple whole-system simulations. 
These simulations are executed entirely with simulated users, referred to as \textit{bots} in the rest of this paper.

The WW bots operate on the real platform infrastructure while being isolated from the production environment. 
This gives us a safe way to explore platform behaviours, in real and counterfactual scenarios.
WW simulations have highly realistic behavioural outcomes, due to the bot interactions being executed on the real platform infrastructure.
More details on Meta's simulation-based approach to analysis, testing and optimisation can be found elsewhere \cite{jaetal:ease-keynote,ahlgren-etal:wes,jaetal:raise-keynote,mhetal:ssbse18-keynote}. 

End-to-End (E2E) Integrity testing forms an important use case for WW simulation.
It is implemented through our MIA simulation-based testing platform~\cite{jaetal:mia}, which runs on top of WW. 
Integrity tests are designed to test the infrastructure for protecting the integrity (also referred to as online safety) of  user communities on Meta platforms. They test the integration of the integrity infrastructure with the overall platform offerings, such as Facebook and Instagram \blue{and are primarily aimed at revealing regression problems due to changes in the system or test infrastructure}.

Integrity tests need to include realistic yet simulated scenarios, such as when a user posts content or behaves in other ways that violate community standards. Testing integrity systems is inherently challenging because such integrity assurance systems incorporate several subsystems that are connected via complex dependencies and precedence constraints, operating to tight operational limitations, and at tremendous scales. 
Simulation is a natural approach to testing such systems. 

However, it comes with its own challenges: any simulation framework will tend to be stochastic which, if not appropriately managed, can cause test flakiness~\cite{mhpoh:scam18-keynote}. Further, for simulation based integrity testing to be effective, we need to ensure that it covers a large space of harmful behaviours and content. 
We refer to this as \textit{test coverage} in the rest of the paper. 

\blue{
The highly varied nature of threats to  user safety involve all forms of content and interaction between uses. 
For example, bugs in the back-end databases, content selection systems or in the logging services can each impact the ability to detect harmful behaviour.  
Although neither back-end database systems, nor content selection systems, nor logging services are directly concerned with integrity, all three feed into the ability of integrity systems to detect and intervene on harmful content and behaviour. 
This means that our tests have to cover executions over the full stack, consisting  of hundreds of millions of lines of code. }

\textbf{Contributions:}
The primary contributions of this paper are:
\begin{enumerate}
    \item We introduce a novel testing tool, \toolname, describing how it uses automated generation of tests from system logs over the Facebook platform and the MIA test platform. \blue{This work reports on industrial experience in the application of inferential test data generation at scale, rather than introducing specific new techniques.}
    
    \item From a system production log space with over 3 million data points, \toolname constructed  2,464 initial test candidates. \toolname further sampled these candidates periodically, filtering them down to produce a refined (highly non-flaky) end-to-end test suite of 39 tests that enjoys a 99.84\% average test pass rate.
    
    \item \toolname creates a pool of test candidates able to  \blue{raise average end-to-end integrity coverage \purple{of one aspect of a particular system} to 24.1\% from a previous (purely manual test design) baseline average of 3.5\% over all violation types}.  Additionally, \toolname increased the \blue{average joint coverage (cross product) of content and violation types from 0.6\% (with manual testing) to 5.4\%}. 
 \purple{This coverage only represents one tool taken into consideration within this paper. There are multiple systems we use to test integrity for which manual testing is the industry standard.}
\end{enumerate}
All results reported in this paper have been obtained from the first three months of deployment of \toolname at Meta, between June and August 2022.

Although the present paper focuses exclusively on the applicability of \toolname to integrity testing, the underlying principles apply more widely to the inference of test signals from checks performed in production systems. Such feedback signals aid in making  systems more robust, leveraging cyber-cyber digital-twins \cite{jaetal:ease-keynote} for software improvement and closing the production-simulation loop.  

\section{End-to-end Integrity Test Automation Approach}
Integrity tests can be categorized into two principle types: \textit{proactive} and \textit{reactive}. 
Proactive tests assess the ability of the integrity infrastructure  to automatically catch violations (e.g., using data-driven classifiers  for automated identification of violating content).

By contrast, reactive tests explore the ability of systems to respond efficiently and effectively to user identification of such content (through reports they make when encountering content or behavior they believe to be in violation of community standards). 
Both kinds of test are important for Meta.
All such forms of integrity test (and many more) are orchestrated using our MIA test system \cite{jaetal:mia}.
Prior to \toolname, all tests were manually written. 
That is, MIA automates test execution and the integration of test signal 
into Meta's continuous integration system, albeit, the tests being designed by engineers.

By contrast, \toolname fully automates test design (inputs and corresponding test oracles) for reactive testing.
\toolname uses MIA to execute the test case, but infers the test steps (actions and corresponding assertions) automatically from production logs.
Specifically, in this paper, we report on the use of a single \toolname test template, in which a bot posts content on a simulated counterpart of the social media platform, such as Facebook or Instagram. 
Another bot reports this content using a set of specific \textit{report tags} (for instance ``hate speech'').
As we get data from production logs, this bot simulates the actions of a real-user that reports content. 

The bot's report enters the same internal reviewing pipeline as used by real users.
This is critical, as the foundational principle of a simulation-based testing approach is that we simulate the users' behaviour, but behaviour is executed on the {\em real} platform infrastructure \cite{jaetal:ease-keynote}.
The content review tool may need to present the reported content to a human reviewer, who then makes the ultimate decisions and invokes the systems that execute the required enforcement/rectification actions.

We also simulate this human reviewer using another bot, since this is a key part of the overall workflow.
It is only human behaviour that is simulated; all interactions between bots are executed on the real infrastructure.
This infrastructural realism ensures low false positive rates \cite{mhetal:ssbse18-keynote}, thereby making any potential bugs identified highly actionable \cite{jaetal:mia}.

The decisions made on the content characterize whether the content is found to be violating and also identify the specific type of violation, for instance ``delete alcohol''. 
This consequently leads to integrity actions being executed, such as the user being checkpointed and/or the content being deleted. 
We refer to this pipeline as \textit{user-report flow}.

\textbf{Problems with manual test design:} 
Manual test design, although the industry norm, is expensive, tedious and slow, and we therefore seek to automate test design wherever possible~\cite{mhetal:ssbse18-keynote,mhpoh:scam18-keynote}. 
Relying purely on manual testing has been known to be sub-optimal, in general, for many years \cite{myers:testing}, and for many reasons all of which also pertain to integrity testing. 
In particular:
\begin{enumerate}
    \item {\bf Familiarity}: Engineers may lack familiarity with the test framework, and test design language. 
    \item {\bf Scale}: Writing tests manually does not scale well with the number of integrity systems that are part of platforms such as Facebook and Instagram, nor does it scale to the rapidly changing integrity threat posture.
    \item {\bf Effort}: Manual integrity test design requires a great deal of time from subject matter experts, which may differ across integrity threats.
    \item {\bf Bias}: As with all human-designed tests \cite{wegener:comparison-journal}, humans have biases and `blind spots' that may cause them to miss important corner cases.
\end{enumerate}

Automated test inference seeks to tackle these problems.
Furthermore, reducing the time to create new tests for integrity systems, and consequently bug identification, are also key desiderata from testing systems for software engineers and subject matter experts. 

\toolname leverages test templates to create, and periodically execute, inferred tests. 
It facilitates continuous and automated testing of new features and code changes. 
Running these tests on MIA ensures a significant reduction in testing time.
For this, \toolname leverages the WW discrete-event based simulation model that emulates user interaction through bots~\cite{jaetal:ease-keynote}.

\section{The Architecture of \toolname}
\label{sec:toolname-architecture}

\toolname is an end-to-end automated testing tool deployed as part of the MIA framework.
\toolname automatically generates user-report flow type tests from logs of human-user interactions over the Meta platforms, such as Facebook and Instagram. 
This entails a completely autonomous process without any human intervention other than generating a generic testing template. 
We assume that human reviewers are accurate in identifying violating content, on which the correctness of such extracted tests relies.  
To do this, we develop an integrity testing template for user-report flows. 
The template we describe below is a canonical example from which we generated all results in this paper. This demonstrates the generic generative power of templates. \blue{A summary of notations is given in Table~\ref{tab:symbols}.}

\subsection{Integrity User-Report Template}
\label{sec:template}

To capture the user-report flow, we identify a generic template that takes a 5-ary tuple as input that we denote by $(t, c, r, d, a)$. Here, $t$ denotes the datestamp when the datapoint was logged in the production databases. $c$ denotes the \textit{content type} for the content to be posted by a bot, and includes some of the types of content that potentially might be considered for actions, for instance ``status update'', ``photo'' and ``comments''. $r$ denotes the list of report tags used by another user to report $c$ (for example, ``bullying'', ``spam'', etc). 

Such a report creates a job for the review tool to which the response is a decision string $d$ (for example, ``delete nudity''). 
Responding to the review job, as per $d$, translates to specific actions being executed on the bot that posted the content and/or the content itself, the list of which is denoted by $a$. 
This may include ``delete''-ing the content or ``checkpoint''-ing the offending user. 
Although there may be multiple calls to such review tools, our template (with its single call to the reviewing pipeline) covers a large fraction (80\%) of production scenarios for the month of July 2022.  

We denote the set of all possible input tuples by $\mathds{I}$. 
We define the template using the notation introduced by Harman et al.~\cite{harman2013comprehensive}. 
We identify two types of activities in tests, \textit{stimuli} and \textit{observations} that are represented by underline $\underline{x}$ and bar $\overline{x}$, respectively.

The template is a function that transforms any input $(t, c, r, d, a)$ into a test activity and a definite oracle that, given a test activity, identifies whether it is acceptable or not. 
We denote a test activity as $\sigma$ and the space of all test activities by $\mathds{S}$. 
We denote a definite oracle by $o$, and the space of all oracles by $\mathds{D}$.  
Now the template is a function $f: \mathds{I} \rightarrow \mathds{S} \times \mathds{D}$ as follows:
\begin{enumerate}
    \item \blue{$\underline{x_1}$: a user bot posts a content of type $c$.}
    \item \blue{$\underline{x_2}$: a user bot reports $\underline{x_1}$ using report tags $r$.}
    \item \blue{$\overline{x_3}$: an review job is created by the system under test.}
    \item \blue{$\underline{x_4}$: a reviewer responds to $\overline{x_3}$ using decision string $d$.}
    \item \blue{$\overline{x_5}$: actions $a$ executed as a response by the reviewer.}
\end{enumerate}
Now, the template works as
\begin{equation}
    (\sigma, o) = f(t, c, r, d, a),
\end{equation}
such that
\begin{gather}
\blue{    \sigma = \underline{x_1} \frown \underline{x_2} \frown \overline{x_3} \frown \underline{x_4} \frown \overline{x_5}},\\
\blue{    o(\sigma) = \overline{x_3} \text{ exists } \wedge \overline{x_5} \text{ exists} \wedge \overline{x_5} = a,}
\end{gather}
\blue{for a previously seen log with $a$ being executed for $\underline{x_1} \frown \underline{x_2} \frown \overline{x_3} \frown \underline{x_4}$ in the production system. } 

\subsection{Generating Tests From Production Logs}
\toolname relies on
the presence of production data that is a collection of  datapoints $(t, c, r, d, a)$. We denote this collection by $\mathcal{P}$. For large-scale platforms such as Facebook and Instagram, running tests for all generated datapoints (in the order of millions) is infeasible within the typical compute budgets of software teams. Thus, we use a sampling function, $s$, that samples a set of datapoints (typically in the order of thousands) from the datapoints logged daily in the production databases. For a given datestamp $t$, we denote the set of sampled tests by:
\begin{equation}
    \omega(t) = s(\{(t', c, r, d, a) | t\ = t \wedge (t', c, r, d, a) \in \mathcal{P}\}).
\end{equation}
The generated tests for a given datastamp $t$ then become:
\begin{equation}
    \{ f(t', c, r, d, a) | (t', c, r, d, a) \in \omega(t) \}.
\end{equation}

However, not all tests in this set may be reliable for continuous use, due to test flakiness.  
Thus, \toolname also filters the initial candidates to produce a suite of
Highly non-flaky tests that enjoy a high pass rate.
This gives us a reusable end-to-end integrity test suite as a final outcome of the inference process.

\begin{table}[t]
    \centering
    \caption{Table of Notations}
    \label{tab:symbols}
    \begin{tabular}{@{}ll@{}}
    \toprule 
    Symbol & Meaning\tabularnewline
    \midrule
    $c$ & Content type \tabularnewline
    $d$ & Decision string \tabularnewline
    $r$ & Report tags \tabularnewline
    $a$ & Integrity actions \tabularnewline
    $\mathcal{P}_{WW}$ & All WW logs \tabularnewline
    $\mathcal{P}_{WW}^t$ & All WW logs collected on datestamp $t$ \tabularnewline
    \bottomrule
    \end{tabular}
    \vspace{-10pt}
\end{table}

\section{The deployment of \toolname at Meta}
\label{sec:toolname-deployment}
The overall orchestration pipeline for \toolname is divided into three phases: exploration, staging and deployment:

\subsection{Exploration Phase}
\label{sec:exploration_phase}

Deploying tests to generate feedback signals is an essential component of an automated-testing pipeline. This must ensure that we not only cover a large-space of tests in production, but also efficiently utilize the computational resources of the software teams. To achieve this we propose three methods that we describe next.

\subsubsection{Optimizing Violation Type Coverage}
\label{sec:violation_type_coverage}
In integrity systems, a key objective is to ensure that we cover diverse types of violations reported in production. Violation types are broad terms that encapsulate several possible sets of report tags. For instance, report tags ``nudity'' and ``nudity sexual activity'' correspond to the violation type ``Pornography''. To ensure we can increase coverage, we need to make \textit{stateful} decisions wherein we capture the tests that have been run for a specific datestamp. 

Specifically, by stateful decisions we mean to create tests informed on the tests generated for the same day, such that we can choose tests not covered in that day and improve coverage. We denote this WW log data by $\mathcal{P}_{WW}^t$ for a datestamp $t$. We represent all WW logs by $\mathcal{P}_{WW}$. For each test generated automatically by the pipeline at date $t$, this database includes all 5-tuple datapoints used to generate these tests. 

We rely on the presence of a function $v$ that converts given report tags (say $r$) to violation types. In order to maximize the coverage of violation types, we generate a sampling score that is defined as follows
\begin{equation}
\resizebox{\hsize}{!}{$\phi_1(t, c, r, d, a) = 1 - \frac{\| \{ (t_i, c_i, r_i, d_i, a_i) \in \mathcal{P}_{WW}^t | v(r_i) = v(r) \} \|}{\underset{v_j}{\max} \| \{ (t_i, c_i, r_i, d_i, a_i) \in \mathcal{P}_{WW}^t | v(r_i) = v_j \} \} }.$}
\end{equation}

The fraction represents the normalized frequency of the violation type corresponding to the input report tags $r$ run in WW at datestamp $t$. Thus, the output of this scoring function is high for violation types with low frequency and vice versa, ensuring the highest sampling score is given to violation types with the fewest runs. 
Consequently, using this scoring criterion makes sure that the daily coverage of violation types is maximized.

\subsubsection{Optimizing Cross Product of Violation and Content Types}
Another critical objective of integrity tests is to ensure that we cover the diverse content types in production logs. However, this must be conditioned to the specific violation types of the corresponding content. For instance, testing all content types, but only for a specific violation type, may not cover the diverse types of rules within the review system. Thus, we consider the cross-product of content and violation types. 
We can leverage a similar strategy as described in Section~\ref{sec:violation_type_coverage} by measuring the run count for each ordered pair of content and violation types.  

Typically, the size of this cross-product space may be too large to distinguish samples in case of limited datapoint availability over the day in question. 
We have indeed observed this in practice for the month of July 2022 in a controlled computational resource setting. Thus, we perform categorical unsupervised clustering to generate run frequencies for a lower number of groups. 

To do this, we use K-Modes clustering algorithm~\cite{chaturvedi2001k}, which is a version of the K-Means clustering approach for categorical data. 
This approach aims to find cluster centers such that all ordered pairs $(c, v(r))$ are assigned cluster centers with the lowest edit distance.  
We denote the function that identifies the cluster index for each input pair $(c, v(r))$ for a datestamp $t$ by $\kappa(c, v(r); \theta)$, where $\theta$ denotes the set of cluster centers learnt in a data-driven fashion. We identify cluster centers $\theta$ by performing clustering of the production data $\omega(T)$. Using this, we identify the scoring function using the run count for each cluster in WW, \textit{i.e.}, $\mathcal{P}^t_{WW}$. Thus,
\begin{align}
\begin{split}
    &\phi_2(t, c, r, d, a) = 1 - \\ & \resizebox{\hsize}{!}{$\frac{\| \{ (t_i, c_i, r_i, d_i, a_i) \in \mathcal{P}^t_{WW} | \kappa(c, v(r); \theta) = \kappa(c_i, v(r_i); \theta) \} \|}{ \underset{(c_j, v(r_j))}{\max} \| \{ (t_i, c_i, r_i, d_i, a_i) \in \mathcal{P}^t_{WW} | \kappa(c, v(r); \theta) = \kappa(c_j, v(r_j); \theta) \} \|}$}
\end{split}
\end{align}
The fraction represents the normalized frequency of the cluster corresponding to the ordered pair $(c, v(r))$ in the WW log data. Thus, the score is higher for those clusters with low frequency in WW logs. This scoring function ensures that both violation and content types are considered when sampling datapoints to generate new tests.


\textbf{Setting hyper-parameters:} A benefit of the clustering-based approach is that the number of clusters is defined by a hyper-parameter, \textit{i.e.}, the number of cluster centers (denoted by $k$), and can be explicitly set. A higher value of $k$ increases the compute time for clustering, whereas a low $k$ reduces the granularity of classification essential to generate sampling scores and disambiguate among datapoints. 
We set $k$ using the \textit{elbow-method}.


\subsubsection{Detecting Anomalies in Production Logs}
The above two methods assume that the production data is deterministic. 
This may not be true at all times due to the inherent stochasticity in mechanisms that rely on human-annotated decision strings to generate the list of actions to execute. 
Social media platforms are a kind of cyber-physical system \cite{jaetal:ease-keynote}, and share similar uncertainty properties with other more traditional cyber-physical systems \cite{zhang:understanding}.

To ensure we generate tests that can facilitate bug discovery, it is critical to cater for such uncertainty and non-determinism. 
To take account of deviations from normal trends in case of integrity actions, we define an outlier score for each datapoint by considering data from two consecutive days:
\begin{align}
\begin{split}
    \phi_3(t, &c, r, d, a) = \mathds{1}(\exists (t_i, c_i, r_i, d_i, a_i) \in \mathcal{P}_{WW} \\ &| t_i = t - 1 \wedge c = c_i \wedge r = r_i \wedge d = d_i \wedge a != a_i),
\end{split}
\end{align}
where $\mathds{1}()$ denotes the indicator function. This scoring function gives a score of 1 to datapoints for which there exists another datapoint in the previous day that matches the input datapoint except for the action list. Thus, sampling using this scoring scheme allows us to pick outliers and detect bugs in the integrity or logging systems.

\subsubsection{Sampling Process}
Now that we have three distinct scoring functions, we build a hybrid sampling approach that considers the three generated scores in tandem. To do this, we take a convex combination of the three scores as
\begin{equation}
\label{eq:covex_combination}
    \phi(x) = \alpha \cdot \phi_1(x) + \beta \cdot \phi_2(x) + \gamma \cdot \phi_3(x),
\end{equation}
where $\alpha + \beta + \gamma = 1$ are hyper-parameters. Here, $x = (t, c, r, d, a)$ is a datapoint in the production logs $\mathcal{P}_{WW}$. 
Without any domain expertise, the default setting is to have equal weights given to the three scores.
We used this default setting to generate the results in this paper. 
We generate the sampling score using Equation~\eqref{eq:covex_combination} and sample $N$ tests to run concurrently with the highest sampling scores, breaking ties uniformly at random. All results in paper are generated with $N = 100$.

\subsection{Staging Phase}
In order to filter out non-production-ready test, \toolname leverages the proportion of passing tests in WW.
This proportion provides an indication of whether these tests are flaky or otherwise unreliable. 
Thus, for each datapoint in WW logs $\mathcal{P}^t_{WW}$, we define two metrics: run count ($\eta$) and pass rate ($\rho$). Formally, 
\begin{align}
    \eta(x) &= \| \{ x \in \mathcal{P}^t_{WW} \} \|, \\
    \rho(x) &= \frac{ \underset{x \in \mathcal{P}^t_{WW}}{\sum} \mathds{D}(x) }{\eta(x)},
\end{align}
where $x = (t, c, r, d, a)$ in the WW logs $\mathcal{P}^t_{WW}$. Note that the denominator in the definition of $\rho$ cannot be less than 1 for any datapoint logged at least once in $\mathcal{P}^t_{WW}$. We define a \textit{tests of interest} for a datestamp $t$, say $x$ as 
\begin{equation}
    x \in \mathcal{P}_{WW}^t \wedge \eta(x) > n_s \wedge \rho(x) > p_s,
\end{equation}
where $n_s$ and $p_s$ are hyper-parameters that represent the thresholds of test runs and pass rate above which a datapoint enters the staging phase. All results in this paper are reported with $n_s = 10$ and $p_s = 0.90$. Using this definition, we periodically run all tests of interest from the WW logs. This enables us to increase the run count of these tests and calculate their pass rate with a higher number of execution samples.

\subsection{Deployment Phase}
We define \textit{production-ready tests} as those that are constructed from a datapoint,  $x$, such that
\begin{equation}
    x \in \mathcal{P}_{WW} \wedge \eta(x) > n_d \wedge \rho(x) > p_d,
\end{equation}
where $n_d$ and $p_d$ are hyper-parameters that represent the thresholds of test runs and pass rate above which a datapoint enters the deployment phase. All results in this paper are reported with $n_d = 50$ and $p_d = 0.95$.  We commit \textit{production-ready tests} as standalone testing functions in MIA and enable alerts to generate signals when any such test fails. 
As these tests are empirically ensured to have low failure rate, we conjecture that any consistent failures after deployment are sufficiently highly likely to be caused by erroneous changes to the integrity production systems.

To automate the process of committing production ready tests in MIA, we use a bot that periodically filters such tests from WW logs and creates production ready code to generate a \textit{diff} to be reviewed by an engineer at Meta (a code change request is referred to as a `diff'). 
\blue{The diff generated by the overall testing workflow is created using exactly the same continuous integration workflow (through Phabricator \cite{feitelson:deployment}) as any regular diff that could have been created by an engineer.
It is submitted for code review, and it's entirely human readable, editable and maintainable.}
In this way, the tool acts as a Practical Automated Tester (PAT) \cite{kbetal:esem21-keynote}.
All deployed tests are removed from the exploration phase to free up computational resources.

\subsection{Example of inferred Integrity Test}

We now present a sample production-ready test deployed in MIA. We first show the datapoint corresponding to the test:
\begin{align*}
    c &= \text{`LiveVideo'},\\
    r &= \text{[`unauthorized sales']},\\
    d &= \text{`delete'},\\
    a &= \text{[`Delete']}.
\end{align*}
This translates to a test activity where a bot shares a `LiveVideo', which is reported by another bot with report tag of `unauthorized sales'. 
This creates an review job that is responded with a decision `delete' and it is asserted that the video is deleted. 
If this test fails, \textit{i.e.}, the assertion fails, because the video is not deleted, then this generates a signal to inform engineering team(s) to check the review rules. 
This also helps us to ensure integrity of the production systems and minimize  harmful content on Meta's platforms.

\section{Evaluation}

In this section, we provide the scientific evidence that the technique improves coverage, finds faults and is able to generate production-ready tests. 
Our evaluation data was collected during deployment in the period between 1 May 2022 and 18 August 2022.
There are two primary desired elements of any automated testing technology.
Specifically, the tool needs to achieve good coverage such that passing tests can reliably translate to there being reasonable confidence in the correctness of the codebase.
The tool should be able to find faults, where there are faults.

\subsection{Coverage}
\label{sec:eval_coverage}

\begin{figure}[t]
    \centering
    \subfigure[Violation Type Coverage]{
    \includegraphics[width=\linewidth]{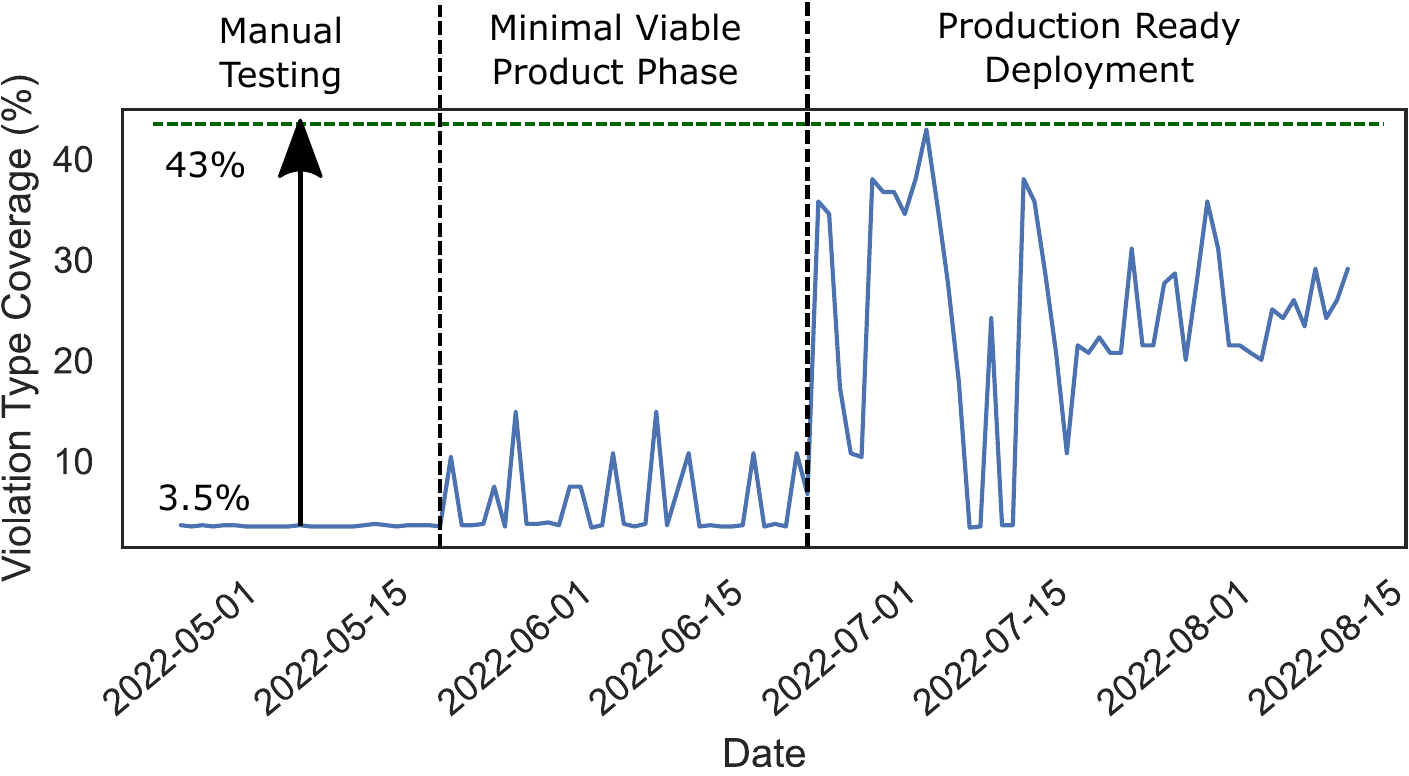}
    \label{fig:violation_type_coverage}
    }
    \subfigure[Cross-Product Coverage]{
    \includegraphics[width=\linewidth]{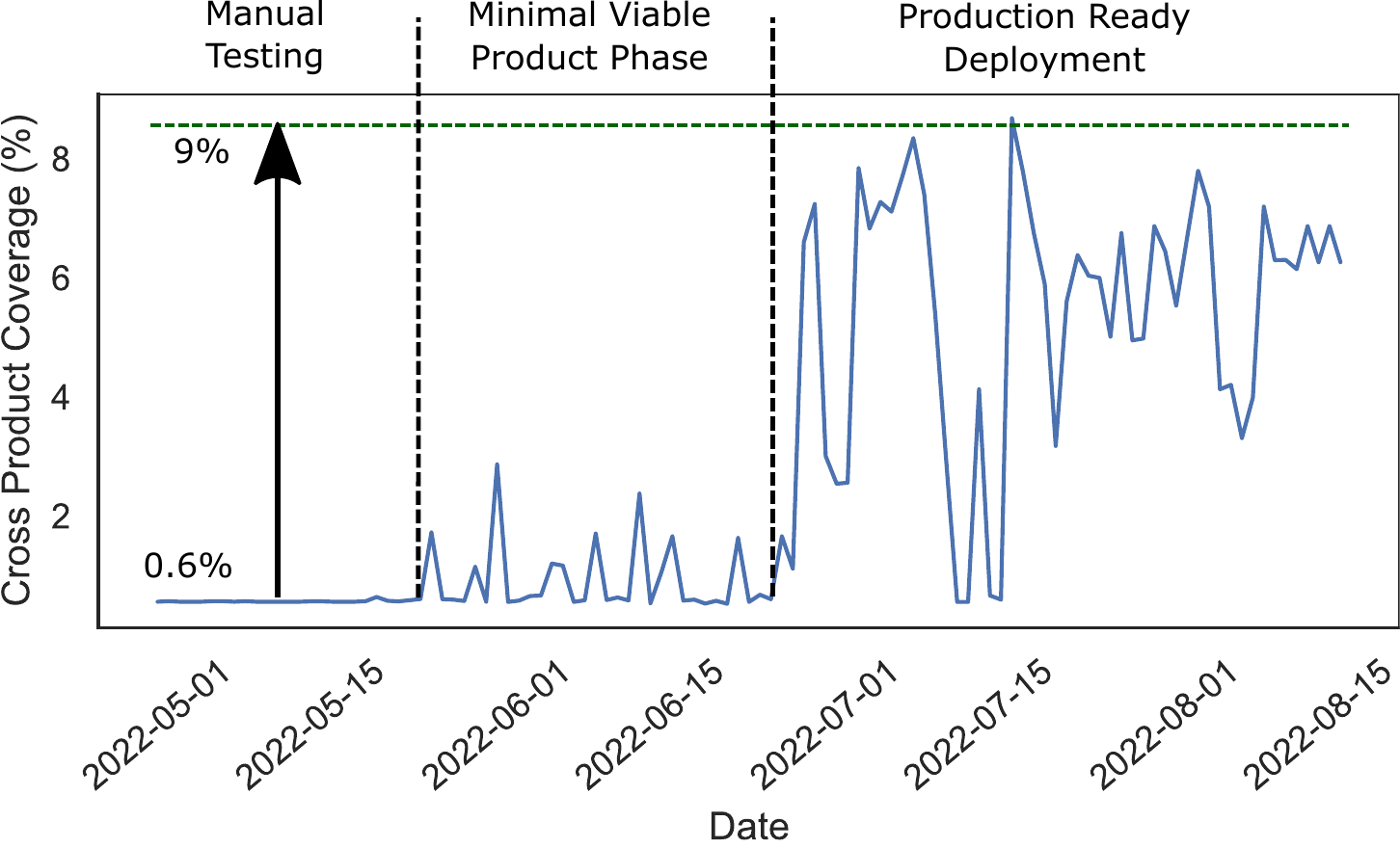}
    \label{fig:cross_product_coverage}
    }
    \caption{Improvement in coverage metrics from manual testing to automated testing \purple{for the particular aspect of e2e testing \toolname was applied to. This coverage only represents one particular tool described in the paper.} We divide the timeline into three phases: manual testing, deployment of a minimal viable product form of \toolname (20 May 2022 to 1 July 2022), and production-ready deployment (1st. July 2022 onwards). The figures demonstrate the improvement in violation type and cross-product coverage by \blue{7$\times$ and 9$\times$}, respectively. The dips in the coverage scores around 15 July are due to identification of bugs wherein the system was temporarily disabled.}
    \label{fig:coverage}
\end{figure}

We seek to answer the following research questions about coverage improvements achieved by \toolname. {\bf RQ1}: what is the improvement in coverage over manual testing ? We divide this research question into two sub questions.
\begin{itemize}
\item {\bf RQ1.1}: what is the improvement in violation type coverage over manual testing?
\item {\bf RQ1.2}: what is the improvement in violation and content type coverage over manual testing
\end{itemize}

To quantify improvement in coverage, we define two metrics (i) violation type coverage, and (ii) cross-production coverage. Both are defined for a specific datestamp and use the data in WW and production logs. 
\begin{align}
    \mu(t) &= \frac{\| \{ v(r_i) \forall (t_i, c_i, r_i, d_i, a_i) \in \mathcal{P}^t_{WW}  \} \|}{\| \{ v(r_i) \forall (t_i, c_i, r_i, d_i, a_i) \in \mathcal{P}_{WW} \} \|},\\
    \nu(t) &= \frac{\| \{ (c, v(r_i)) \forall (t_i, c_i, r_i, d_i, a_i) \in \mathcal{P}^t_{WW}  \} \|}{\| \{ (c, v(r_i)) \forall (t_i, c_i, r_i, d_i, a_i) \in \mathcal{P}_{WW} \} \|}.
\end{align}
The former is the fraction of violation types covered in the WW logs with respect to those in production data. The latter is the fraction of the cross-product of content and violation types covered in WW with respect to the ones in production.

To test this, we deploy \toolname in a phased manner, \textit{i.e.}, first deploying a minimal viable product only with the exploration phase on 20 May, and then deploying the complete pipeline on 1 July. We run \toolname in a controlled compute environment with a capacity akin to the typical budgets of software development teams. Figure~\ref{fig:coverage} shows the improvement in violation type and cross-product coverage. 
\blue{On an average, the former increases from 3.5\% to 24.1\% (a 7$\times$ improvement) and the latter from 0.6\% to 5.4\% (a 9$\times$ increase). The highest observed violation type and cross product coverage are 43\% and 9\% respectively.}

This answers RQ1: when compared to existing manual testing techniques, \toolname achieves an average \blue{7$\times$} improvement in violation type coverage and \blue{9$\times$} improvement in coverage of the cross product violation and content type.
Overall, therefore, we find evidence that automated test inference using \toolname has achieved \blue{dramatic improvement in test coverage}, when compared to the purely manual testing that preceded it.

\subsection{Variation in Distributions}
Any automated testing approach needs to produce non-flaky tests.
End-to-end test such as this are widely known to be subject to high degrees of flakiness, as has been reported previously by several organisations \cite{luo:flaky,mhpoh:scam18-keynote,memon:taming}.
Fortunately, since inferred test are generated from production data, scale is more of a opportunity than a challenge; the more production data is available, the more candidate test can be made, and therefore the higher we can set the bar for non-flakiness of generated tests.

A critical research question we seek to answer concerns the degree to which \toolname is able to filter out candidate tests to remove those that exhibit even small amounts of flakiness, to yield a highly non-flaky, and thereby a highly actionable, inferred test suite.
We therefore report on the behaviour of our approach over the three stages of the \toolname pipeline, \textit{i.e.}, exploration, staging and deployment. We seek to answer: \textbf{RQ2:} How does the inference pipeline filter inferred tests? We decompose this into two sub-questions:
\begin{itemize}
    \item \textbf{RQ2.1}: how does the number of tests and test pass rate vary across three stages of the inference pipeline?
    \item \textbf{RQ2.2}: how does the distribution of content and violation types change at each stage of the test inference pipeline?
\end{itemize}

\begin{table}[t]
    \centering
    \caption{Summary of the number of tests and the average pass rate for tests in each phase.}
    \begin{tabular}{@{}cccc@{}}
    \toprule
        Phase & Exploration & Staging & Deployment \\
        \midrule
        Number of tests & 2464 & 42 & 39 \\
        Average pass rate & 0.307 & 0.9971 & 0.9984\\
        \bottomrule
    \end{tabular}
    \label{tab:phase_summary} \vspace{-10pt}
\end{table}

\begin{figure*}
    \centering
    \includegraphics[width=\linewidth]{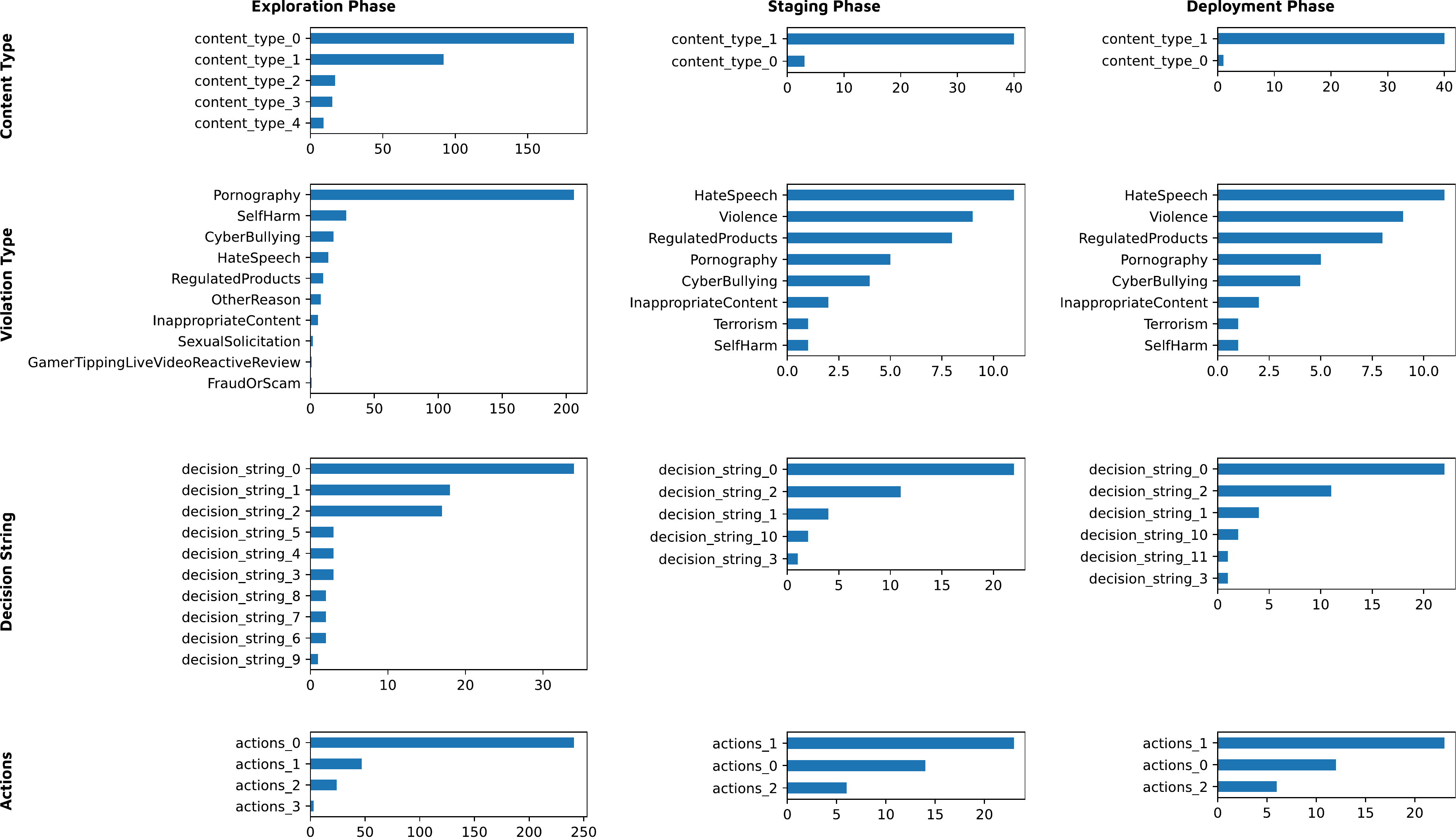}
    \caption{Distributions of content type, violation type, decision string and actions in the exploration, staging and deployment phases. Corporate sensitive information not relevant in justifying the scientific claims have been replaced with identifiers to compare relative changes in distributions across the stages of the pipeline. This includes the decision strings and actions.}
    \label{fig:distributions}
\end{figure*}


Table~\ref{tab:phase_summary} shows the number of tests as well as their average pass rates in each phase. As the tests progress from exploration to staging to deployment, the average number of tests in each phase drops. As the conditions to filter tests in subsequent phases become stricter, the average pass rate increases, answering RQ2.1. Distributions of the content types, violation types, decision strings and actions for each phase are shown in Figure~\ref{fig:distributions}. The exploration phase covers more diverse types of contents, violation types, decision strings and actions, thanks to the intelligent sampling procedure in this phase that aims to maximize not only coverage but also select outliers. In contrast, the specific content or violation types that have the highest pass rate are more common among the tests in the staging or deployment phase, answering RQ2.2. 

It is interesting to observe how the content types and decision strings covered follow and approximate power law: some types of content, decision string and action have a higher cover, with an apparent long tail of other content types, actions and decisions strings. 
We conjecture that this will be experienced in other similar inferential testing scenarios as well.
That is, our experience indicates that it is inherent in the process of inferring test from production logs, that there will be non-uniform coverage.
Indeed, we observe that the distribution can be highly non-uniform, due to the combined effects of the availability of suitable production data, and the anti-flaky filtration process.

The results for the distribution of content type do not necessarily reflect those on the platforms tested as a whole, but rather the ability to infer reusable tests from production logs relating to these content types.
Although the technical details will differ from application to application, there will be non-uniform distributions of available salient features of inferred tests, no matter the domain to which inferential testing is applied.

With this in mind, we believe that one of the contributions of our work is the intelligent sampling procedure.
This serves to ensure all content types, and corresponding decision and action types are considered. 
Without this, inherent non-uniform distributions, such as these, may lead to some aspects going untested.
This is particularly important since the filtration process may also disproportionately remove certain tests.
For example, candidate tests that happen to cover aspects of the system that are highly nondeterministic will exhibit a higher degree of flakiness, thereby increasing the risk that they are removed disproportionately and discarded by filtration.
In order to ensure good coverage of the overall test suite, it is important that we maximise selection opportunities by presenting the widest diversity of candidate tests at the commencement of the filtration process.

\subsection{Surfacing of Bugs}

The end-requirement of any testing process is that it should be able to surface bugs. 
From a scientific point of view, it is useful to understand the kinds of {\em engineering} issue that can be tackled using a technique like inferential test  construction.
Therefore, in this section, we give a characterisation of the engineering properties of the kinds of bugs that we are able to find using our inference technique,  \toolname.
However, because our testing enhances online safety, we obviously must refrain from revealing any details that may be useful to those with harmful intentions.

Like any approach to software testing, a considerable amount of infrastructure needs to be built in order to capture and execute tests. We essentially have three systems working together to achieve this inferential testing approach.
That is, we have the integrity system under test, which is concerned mainly with protecting against, reporting and acting on malicious behaviours on the platform. 
We have the test infrastructure, MIA, which is used to automate the process of test execution and which acts as a platform for end-to-end simulation-based testing of the whole platform (including the specific integrity systems tested).
Finally, we have the inference engine and supporting infrastructure for capturing the system data required to implement the inferred tests; the \toolname code itself.

Execution of the inferential test suites was able to find issues in all three of these key components. Naturally, any approach to developing test cases will find bugs in itself, but this is uninteresting from a Software Engineering point of view; software developers routinely find issues as they develop and so there is nothing surprising or unusual in this scenario.
However, perhaps more interesting, from an engineering standpoint, is the way in which the deployment of inferential testing helped us to find issues, implicit assumptions, and unexpected constraints, in the underlying test infrastructure itself.
That is, \toolname gives us the ability to scale up testing, to increase coverage, and also to run the kind of tests that human developers would not necessarily think of executing, but which are, nevertheless, useful.

Specifically, the combination of scale with  tests that might not be written as human testers might write them, did reveal issues in our testing infrastructure itself. 
It is unlikely that such issues would have been so easy to discover using purely human-designed tests, but some of these issues may have affected the human-designed tests, over longer time frames, nevertheless.
In this way, inferential testing acted as a kind of traffic-based `stress test' \cite{garousi:traffic,delgrosso:stress} on the test infrastructure, thereby allowing us to quickly identify and address issues that might otherwise have take far longer to uncover and tackle with purely human-designed test cases.

Finally, as one would expect, with any testing approach, we are able to find and fix bugs in the system  under test. 
In this case, from an engineering point of view, the kind of bugs found fall into all of the usual categories that we expect from any general testing systems including, for example,  initialisation, logging, and update issues. 
All of these were identified early and fixed using the approach.

\subsection{Lessons Learnt}

\blue{We now discuss key takeaways from our industrial application of test inference using  {\toolname}:}
\begin{enumerate}
    \item \blue{Inferring oracles is extremely hard, considering the dynamism in the characteristics of the social interactions among humans. Novel interaction types frequently lead to unseen scenarios for which pre-defining oracles is challenging.}
    \item \blue{Within the limited computational budgets, leveraging production logs enables inferred tests to focus on common and critical cases.}
    \item \blue{Filtering tests to prioritise the execution of non-flaky tests generates a dependable signal, allowing engineers to take informed decisions on code changes.}
    \item \blue{The more important bugs found have led to in-depth investigation, by engineers, with meetings to discuss and follow up upon lessons learnt. This provides evidence that inferential test generation techniques can reveal non-trivial bugs in highly complex real-world systems.}
\end{enumerate}

\section{Related work}

There has been a great deal of work on test automation in the last few decades, starting with early pioneering work on automated search-based software testing \cite{mh:icst15-keynote} and symbolic execution \cite{cadar:three-decades}, both of which can be traced by to  1976 \cite{miller76_floating_pt_test_data,boyer:select, clarke:symbolic}, and which have remained active topics of research ever since.

An ever-present problem throughout these five decades of research has been the lack of a test oracle~\cite{ebetal:oracle}, first formally identified and fully described by Weyuker in 1982 \cite{weyuker:untestable}.
The Oracle Problem is concerned with determining, automatically, whether a test has passed: while we can automate the process of generating test inputs, the problem of determining which outputs should be deemed correct for giving input has remained challenging.

When there is a specification of the intended behaviour, then the oracle problem is less pernicious, because it is transformed into the problem of generating oracles from specifications \cite{coppit:use,richardson:specification}.
This approach can be adopted even when the specification is written in a (structured) natural language, rather than a formal specification language \cite{motwani:automatically}.

However, in many cases there is no specification, and it remains impractical to construct one.
Furthermore, for some systems it has proved to be impossible to fully specify the expected behaviour.
Simulations are one such category of systems: we use simulations, in general, to discover, {\it inter alia},  outcomes that may be unexpected.
As such, simulation-based systems denote the epitome of Weyuker's `untestable' programs \cite{weyuker:untestable}: not only is the specification typically not known, but it is inherently {\em unknowable}.
If we could accurately predict the real behaviour we would not need a simulation, 
so we cannot expect to have a fully automated oracle for a simulation-based system.

For regression testing, the output from the previous version of the system can be used as a partial oracle for subsequent executions \cite{arrieta:using}.
However, this leaves open the problem of determining which aspects of previous version's behaviour are salient for the regression test problem in hand.
It also clearly cannot cater for situations in which the change in question seeks to fix previously buggy behaviour. 

Finally, regression testing is only one {\em part} of the overall test obligations that rest on software engineers' shoulders.
Specifically, the oracle problem remains particularly challenging for testing novel functionality, for which previous behaviour is, by definition,  unavailable  and thus regression testing is inapplicable.

Opportunities for extracting test information from production logs have been known for some time \cite{amalfitano:rich}.
Full test automation (test design including inputs and expected output behaviours) is  possible using the so-called implicit oracle~\cite{ebetal:oracle}, which captures known buggy behaviours like crashing and running out of memory. 
Such behaviours can be deemed to be incorrect behaviour, irrespective of the input used to test the system under test. 
At Meta Platforms, we have previously used an implicit oracle in our work on simulation-based testing using Sapienz~\cite{mhetal:ssbse18-keynote} and FAUSTA~\cite{kmetal:fausta} for example.
The implicit oracle is a powerful way to overcome the oracle problem for testing problems associated with reliability, both client-side application reliability (in the case of Sapienz) and server-side reliability (in the case of FAUSTA).

However, the aim of moving beyond the implicit oracle poses considerable challenges; the automated construction of oracles requires knowing the {\em intended} behaviour of a software system as well as its {\em actual} behaviour.
Technologies for automatically searching test input spaces are now relatively mature and widely deployed.
Nevertheless, the problem of automatically determining acceptable outputs remains one of the final obstacles to fully automatic test case design.
As a result, there has been much work on approaches to automated partial oracle discovery,  synthesis and inference~\cite{pezze:automated,oliveira:automated}.

Work has also been conducted on automated inference of test oracles from documentation \cite{peters:using,blasi:memo},
exceptions \cite{goffi:automatic},
and
mutation testing \cite{fraser:mutation-driven,staats:automated}.
Techniques aim to alleviate some of the human burden, have focused on either reducing human effort to define oracles \cite{afshan:evolving,pmetal:stov10} or on providing  better automated support to help humans improve oracles \cite{jahangirova:issta16}.
Another approach to reducing human oracle effort is simply to reduce the number of test inputs automatically generated, such that the human effort of checking the corresponding outputs is also reduced \cite{mhetal:oracle}.

Metamorphic testing \cite{segura:metamorphic-survey} and other `pseudo' oracle approaches \cite{mcminn:co-tetra} seek to reduce oracle effort (manual or otherwise) by capturing only properties or aspects of oracles rather than the complete oracle. 
Metamorphic testing further reduces the cognitive burden on the human test designer, who merely has to define metamorphic relations rather than full test oracles \cite{segura:metamorphic-survey}.
Likely metamorphic relations can also be inferred, thereby more fully automating the metamorphic testing approach \cite{su:dynamic}.

Much of this existing literature on test inference, promising though it is, has hitherto been evaluated on relatively small-scale non-industrial systems and none of the previous literature reports on results from industrial deployment at scale in the field.
By contrast, \toolname has been deployed for integrity testing, at Meta, on systems of hundreds of millions of lines of server-side code. 
To the best of our knowledge, the present paper, therefore, represents the first report of an industrial deployment of automated oracle inference at scale.

\subsection{Relation to Capture--Replay, Regression and Combinatorial testing}
\blue{
ALPACAS could be thought of as closely related to capture--replay test techniques \cite{steven:jrapture} (especially those that rely on execution logs \cite{orso:selective}) because it constructs tests from observations. 
The observations could be thought of us being `captured', and the test is `replaying' these captured observations.}

\blue{Also, like capture--replay techniques, ALPACAS is also a form of regression testing.
However, compared to more general regression test generation methods~\cite{arrieta:using,symh:regression-survey}, ALPACAS does not need explicit supervision on the specific aspects to be tested, nor oracles to be explicitly specified. ALPACAS circumvents this by leveraging the human generated logs to use them as a proxy for the oracle behaviour. }

\blue{Similarly, other works that rely on an implicit oracle~\cite{ebetal:oracle} are biased by the implied conclusions (from crashes or buggy behaviours) or assumptions (about the working of the underpinning systems). Such systems are typically unable to quickly adapt in highly dynamic systems such as those at Meta Platforms. Other oracle inference processes such as explicit human specification are known to not scale well~\cite{mhetal:oracle}. 
}

\blue{Combinatorial testing  \cite{nie:survey,hwetal:empirical-combinatorial} is complementary to ALPACAS, because the $(c, r, d, a)$ values inferred from the human production logs could be used to explore additional interactions.
For example, we could seek to explore pairwise interactions of suitably selected ups are values, using combinatorial testing minimise the number of tests involved.
Thus, a total interaction testing procedure could be used as an extension to ALPACAS.}

\section{Limitations}

The present study demonstrates the applicability of \toolname using a single test template. 
Although we are able to cover a large fraction of cases (80\%) using the template presented in Section~\ref{sec:template}, this does not cover integrity scenarios that have multiple calls to the reviewing pipeline. 

Further, Figure~\ref{fig:coverage} also highlights the high level of stochasticity in the testing inference process. 
This is due to non-determinism in the computational infrastructure and 
\toolname behaviour he's also sensitive to changes in the underpinning infrastructure and the MIA framework.

\toolname is specific to integrity systems, and in particular those deployed on platforms such as Facebook and Instagram for social media.
Nevertheless, we believe that the approach we have adopted, and the observations on which we report in the paper would be useful for other inference approaches, including those for other integrity systems and for systems that do not concern integrity at all.

Clearly, the specific template we use is pertinent to, and specifically constructed for, the test scenarios in hand. 
This would be the case with any test inference approach similar to ours. 
Apart from this necessary scenario-specific detail, the overall approach of inferring tests from system production logs, deploying these, filtering out flaky tests, and observations about distributions are likely to apply in other inferential testing scenarios.
We therefore hope and believe that, despite the necessary limitations of our own approach, the results reported in the present paper will generalise, and will inform future development of inferential testing, research and practice.

\section{Future Work and open problems}

There are several straightforward directions for future work to develop \toolname at Meta:

\noindent {\bf Extending to proactive integrity tests}:
     The current implementation considers automation of only reactive test types. However, the pipeline can be extended to proactive classification based tests. This would allow us to automatically test integrity systems that leverage pre-trained classifiers that take-down violating content on Meta's platforms. Similarly, the tool can be extended to test systems other than those related to integrity that may use production logs different from those generated by bot-interactions. 

\noindent {\bf Extending to root cause identification}:
The current approach can be extended to further investigate and identify root-causes of bugs. Specifically, the identification of code changes that gave rise to bugs may be of value to software engineers to facilitate a reduction in remediation time. 
    This further allows us to identify point-of-contacts to alert or notify for informed remediation of identified bugs.

The work we report in this paper also suggests future directions and open problems:

    \noindent {\bf Extending beyond integrity testing:} While we have applied our techniques to inferring integrity tests, the approach could also be extended to infer tests to tackle other software engineering concerns, such as privacy, reliability, and performance. 
    In general, any software system that is deployed at scale will have a large amount of production data from which tests can be inferred.
    This offers the opportunity to scale, not only the amount of testing that can be done, but also to tackle the hitherto highly  pernicious Oracle problem \cite{ebetal:oracle}.
    
    \noindent {\bf Testing counterfactual scenarios:} In our work, we have inferred tests from production and run them on a simulation-based testing platform. One of the important properties of simulation based testing is its ability to test counterfactual scenarios \cite{jaetal:ease-keynote,jaetal:mia,Ie:RecSym}.
    
    Therefore an interesting open problem is the generalisation of inferred tests to tackle such scenarios.
    For example, suppose we have inferred a test, based on production observations, in which a particular parameter is always set to a specific value. 
    If we now vary this value, this allows us to test how our systems might respond to an unexplored and, as yet, not encountered scenario.
    Searching hitherto unexplored, yet feasible, traffic offers us the possibility to deploy advanced preventative measures against potential future problems.
    
    By varying the inferred test in this way, we are essentially conducting highly realistic tests (even though they concern counterfactual scenarios). 
    The tests retain realism, because they follow large subsections of existing workflows of production traffic that have already experienced, with only small `counterfactual tweaks'.
    The open research challenge here is to identify the values to be permuted in this way in order to test meaningful and actionable counterfactual scenarios, while maximally retaining realism.
    
    \noindent {\bf Inferred test support for Genetic Improvement}: Genetic Improvement \cite{Petke:gisurvey} uses a test suite to guide the search for improvements to the software code (the `genetic' material).
    It seeks to optimise one or more properties, while resisting regressions in functionality that is to be retained. 
    Similar to other search based improvement approaches, including automated program repair \cite{legoues:cacm-survey}, the quality of the improvements, and their ability to resist regression, is crucially determined by the strength of testing.
    
    By inferring test cases we have opportunities to scale testing because we can generate arbitrary numbers of tests.
    Since inferred tests also crucially contain their corresponding oracles, we have the ability to scale the generation of test suites the guide genetic improvement -- at least in principle -- if we can also target test generation at particular areas of code. 
    
    This suggests an interesting open research problem at the intersection between test inference and genetic improvement: How do we target the test inference to particular areas of code that are subject to improvement using test-guided techniques such as genetic improvement and automated program repair?

    \noindent \blue{{\bf Improving the orchestration pipeline}: The overall orchestration pipeline of \toolname uses metrics of violation type coverage, cross product coverage and anomaly detection to sample  tests. Our convex combination based sampling approach is sufficient to demonstrate the impact of inferential testing. However, further analysis of the Pareto front of the optimization problem (and employing a multi-objective approach) is worth exploring in the future.}

\section{Conclusions}
This paper reported on the first deployment of industrial scale, fully automated software testing through test inference from production logs.
We focused on the inference of tests that assess the ability of integrity systems to react to user-reported content violation, but the principles reported in this paper apply more widely.

\blue{Manually writing tests requires familiarity with the underpinning system and human effort. 
ALPACAS tests correspond to actual user reports and all bot interactions are executed in the real Facebook platform, hence all inferred tests are realistic.
This allows for achieving the widest coverage as well as lowest flakiness and false positive rate. }

The tests we infer are executed on our MIA test infrastructure, built on the WW simulation platform. 
This platform simulates user interactions with the integrity system, but not the system itself, thereby yielding highly actionable inferred test suites.
We evaluated our approach, showing that it is able to generate a realistic test suite (39 tests) with a low flakiness and false positive rate  ($>$99\% pass rate). 

Automation is particularly important for integrity, because an entirely human-based approach clearly cannot scale to tackle the testing of large number of interconnected integrity sub-systems. 
However, the approach we adopted and the results on which we report here are only specific to integrity in their implementation details.
Our approach therefore generalises to other automated test inference problems outside of the domain of software safety/integrity.

\noindent{\bf Acknowledgements}:
The authors would like to thank the many Meta engineers, managers and leadership for nurturing, encouraging and helping to develop the work reported here. 
Shreshth Tuli  is supported by the President's PhD Scholarship at Imperial College London.
Mark Harman's scientific work is part supported by European Research Council (ERC), Advanced Fellowship grant number 741278; Evolutionary Program Improvement (EPIC) which is run out of University College London, where he is part time professor. He is a full time Software Engineer at Meta Platforms Inc.

\balance

%

\bibliographystyle{IEEEtran}
\bibliography{slice,sample-base}

\end{document}